\documentclass[11pt]{article}

\usepackage[final]{acl}

\usepackage{times}
\usepackage{latexsym}
\usepackage{booktabs}
\usepackage{tabularx}
\usepackage{tcolorbox}
\usepackage{amssymb}
\usepackage{adjustbox}
\usepackage{algorithm}
\usepackage{algorithmic}
\usepackage{pifont}
\usepackage{xcolor}
\usepackage{listings}
\tcbuselibrary{breakable}

\usepackage[T1]{fontenc}

\usepackage[utf8]{inputenc}
\usepackage{multirow}
\usepackage{microtype}

\usepackage{inconsolata}

\usepackage{graphicx}

\title{Disentangling Structure and Semantics: How Schema Representation Affects LLM-Based SQL Generation.}

\author{
 \textbf{Daniel Yitian Su\textsuperscript{1}},
 \textbf{Sophie Yiran Su\textsuperscript{1}},
 \textbf{Qiang Sun\textsuperscript{1}},
 \textbf{Yihao Ding\textsuperscript{1}} \thanks{Corresponding Author.},
 \textbf{Wei Liu\textsuperscript{1}} \thanks{Corresponding Author.},
\\
 \textsuperscript{1}The University of Western Australia, Crawley, Australia
\\
 \small\textbf{Correspondence:}\texttt{daniel.su@research.uwa.edu.au} \texttt{\{yihao.ding, wei.liu\}@uwa.edu.au}
}

\tcbset{
    colback=gray!10,       
    colframe=gray!20,      
    arc=4pt,               
    boxrule=0.5pt,         
    left=6pt, right=6pt, top=6pt, bottom=6pt, 
    fontupper=\small\ttfamily, 
    halign=left
}

\begin{document}
\maketitle
\begin{abstract}
LLM-based text-to-SQL pipelines read the database schema as text, which carries both \emph{structural} cues (tables, keys, relationships) and \emph{semantic} cues (table and column names); prior work has studied each axis in isolation, leaving open how they compare in magnitude and whether they substitute for one another. We present a controlled $6\times3$ factorial design crossing structural levels $L_1$--$L_6$ (from a denormalised wide table to a 3NF schema with foreign keys and explicit join paths) with semantic levels $S_1$--$S_3$ (anonymous, abbreviated, descriptive identifiers), evaluated on 397 corrected BIRD questions with identical gold queries throughout; we materialise 1NF and 2NF variants for nine BIRD databases to support the lowest structural levels. Across nine models from 0.5B to flagship scale we find an \emph{asymmetric substitution} between the two axes, meaningful names compensate for missing structure but richer structural metadata does not recover performance when names are opaque, which reproduces in 8 of 9 databases and emerges with model scale (negligible below 3B). Within the structural axis the dominant lever is \emph{normalisation itself}, not metadata layered on top of 3NF, suggesting that for current LLM-based text-to-SQL the practical bottleneck is semantic grounding rather than relational exposure.

\end{abstract}
\section{Introduction}
Text-to-SQL, the task of translating natural language questions into executable SQL queries, has emerged as one of the most actively studied interfaces between non-expert users and relational databases \citep{yu2018spider,li2023can}. With the rapid progress of large language models (LLMs), prompt-based LLM pipelines have become the dominant paradigm, surpassing earlier sequence-to-sequence and graph-based parsers and now occupying the top of public leaderboards such as BIRD and Spider \citep{pourreza2023din, gao2024preview, liu2025survey}. In this paradigm, the model rarely sees the underlying data and instead reads the database schema as text in the prompt. This schema text carries two kinds of information: \emph{structural} cues such as tables, keys, and relationships, and \emph{semantic} cues carried by table and column names. How these cues are written can change execution accuracy substantially, even when the model, the question, and the database are held fixed \citep{chang2023prompt, gao2024preview}.

Despite this sensitivity, prior work has examined the two axes only in isolation. On the structural side, \citet{kohita2025exploring} studies how normalization levels affect SQL generation, while on the semantic side, \citet{luoma2025snails} shows that schema naming quality strongly influences accuracy. Earlier comparative prompt studies \citep{chang2023prompt, gao2024preview} vary multiple aspects of the schema text at once, so the resulting accuracy differences cannot be attributed to either axis alone. As a result, three questions remain open: (i) how the two axes compare in magnitude on a shared benchmark, (ii) whether they substitute for or complement each other, and (iii) how these effects behave on a real, large-scale benchmark such as BIRD rather than on single-domain schemas.



To resolve these questions, we run a controlled factorial study that crosses two independent axes of schema representation. The \emph{structural axis} spans six levels (L1-L6), from a single denormalized wide table, through 2NF clusters, to a fully normalized 3NF schema progressively augmented with types, foreign keys, and explicit join paths; the \emph{semantic axis} spans three levels (S1-S3) of identifier informativeness, from opaque names through abbreviations to descriptive English. Crossing the two yields 18 conditions, evaluated on a fixed set of 397 questions from a corrected BIRD mini-dev subset \citep{li2023can} with the same gold queries and execution-accuracy metric, so any difference can be attributed solely to schema representation. Because BIRD provides only 3NF databases, we additionally materialize 1NF and 2NF variants for nine of its databases to support the lowest two structural levels.

Our contributions are: 
i) We present a controlled analysis framework for studying how schema representation affects Text-to-SQL performance, disentangling structural observability (L1-L6) from semantic richness (S1-S3) across 18 conditions over a fixed question set.
ii) We construct materialised 1NF and 2NF variants for nine BIRD databases and provide a reproducible denormalisation pipeline, enabling controlled comparison between normalised and denormalised schema representations.
iii) We find an asymmetric substitution effect between semantic and structural cues: meaningful schema names can compensate for weak structural observability, while richer structural metadata cannot recover performance when names are opaque.
iv) We show that schema normalisation changes the composition of model errors rather than simply reducing them, shifting errors from fan-out and deduplication issues toward join and table-selection failures.

\section{Related Work}
Prompt-based LLM pipelines now dominate Text-to-SQL leaderboards on Spider \citep{yu2018spider} and BIRD \citep{li2023can}, with systems such as DIN-SQL \citep{pourreza2023din}, DAIL-SQL \citep{gao2024dail}, and XiYan-SQL \citep{gao2024preview} relying on the serialised schema as their primary view of the database, and recent surveys \citep{liu2025survey} documenting the shift to this paradigm. \citep{chang2023prompt} systematically vary how the schema is rendered in the prompt, serialisation format, inclusion of sample content, and demonstration design, but co-vary structural and semantic attributes, so the contribution of each cannot be isolated. More recent work targets a single axis. On the structural side, \citep{kohita2025exploring} examines normalisation levels from 1NF to 3NF across eight LLMs, finding a query-type-dependent trade-off: denormalised schemas favour simple retrieval while normalised schemas better handle aggregation. Robustness benchmarks such as Dr.Spider \citep{chang2023drspider} further show that even minor schema perturbations, abbreviating or renaming columns, cause large accuracy drops, implicitly implicating identifier semantics as a fragility source. On the semantic side, SNAILS \citep{luoma2025snails} demonstrates that identifier naturalness has a statistically significant effect on accuracy across multiple LLMs and prompting workflows, while \citep{wretblad2024synthetic} show that augmenting prompts with LLM-generated column descriptions consistently improves performance, with richer metadata helping even when human annotators consider it redundant. Neither line of work compares the two axes on a shared benchmark or analyses their interaction. We address this gap by crossing both axes in a single factorial design over a fixed BIRD-derived question set, with materialised 1NF/2NF variants to support the lowest structural levels.

\section{Methodology}
\label{sec:method}
\subsection{Problem Setup and Overview}
\label{sec:overview}

The text-to-SQL task takes a natural language question $q$, an underlying database $D$, and a schema text $R(S)$ embedded in the prompt, and asks the model to generate an executable SQL query whose result on $D$ matches that of the gold query. We measure correctness with execution accuracy (EX), the standard metric on BIRD and Spider. Our study investigates how $R(S)$ shapes the model's behaviour along two largely independent axes: a \emph{structural} axis that controls how much relational structure is exposed to the model, and a \emph{semantic} axis that controls how informative the schema identifiers are. Each axis is discretised into a small number of levels (defined in §\ref{sec:schema_rep}), and the Cartesian product of the two yields 18 controlled conditions. Across all 18 conditions, the questions, gold queries, EX metric, model, and underlying database content are held constant; only the schema text in the prompt varies. This factorial design isolates the marginal and interaction effects of structure and semantics on generation accuracy. The remainder of this section describes the dataset (§\ref{sec:dataset}), defines both axes (§\ref{sec:schema_rep}), presents the materialisation pipeline used for the lower structural levels (§\ref{sec:denormalisation}), and details the evaluation protocol (§\ref{sec:evaluation_protocol}).

\subsection{Benchmark Setup}
\label{sec:dataset}

We construct the evaluation benchmark in three steps: \textbf{1) source benchmark selection}, where we adopt BIRD because of its large scale, cross-domain coverage, and focus on complex, real-world SQL generation; \textbf{2) gold annotation correction}, where we address the substantial noise in BIRD's original gold SQL annotations, with reported error rates up to 52.8\%~\cite{jin2026pervasive}, by evaluating on Arcwise-Plat-SQL \footnote{\url{https://github.com/uiuc-kang-lab/text_to_sql_benchmarks}}, a corrected BIRD mini-dev subset released by the same authors that revises erroneous gold queries while keeping the original questions and databases unchanged; and \textbf{3) database filtering for materialisation}, where we further restrict the corrected pool to \textbf{nine databases} for which 1NF and 2NF materialisation via chained full outer joins remains computationally tractable. We exclude \texttt{card\_games} and \texttt{codebase\_community} because their dense many-to-many relationships and large schema graphs produce intractable wide-table row counts (see §\ref{sec:denormalisation}). The resulting benchmark, shared by all 18 experimental conditions, consists of:
i) \textbf{397 questions} drawn from the corrected set;
ii) \textbf{9 databases} in their original BIRD SQLite form, used unchanged as the 3NF baseline for the upper structural levels; iii)
materialised \textbf{1NF and 2NF} variants of each database, used by the lower structural levels (constructed in §\ref{sec:denormalisation}).

\subsection{Schema Representation}
\label{sec:schema_rep}
We define the two axes of schema representation that the factorial design varies across. The \emph{structural axis} controls how much of the database's relational structure is exposed to the model (§\ref{sec:structural_axis}), while the \emph{semantic axis} controls how informative the schema identifiers are (§\ref{sec:semantic_axis}). All other elements of the task instruction, question text, output format, remain identical across the 18 conditions.

\subsubsection{Structural Axis}
\label{sec:structural_axis}

\begin{table}[t]
\centering
\caption{Structural levels ($L_1$--$L_6$). DB denotes database form; Type/PK, FK, and Join denote whether column types/primary keys, foreign keys, and explicit join paths are included in the prompt.}
\label{tab:structural_levels}
\small
\setlength{\tabcolsep}{3pt}
\begin{tabular}{clcccc}
\toprule
Level & Description & DB & Type/PK & FK & Join \\
\midrule
$L_1$ & Single 1NF wide table      & 1NF & -- & -- & -- \\
$L_2$ & 2NF wide clusters          & 2NF & -- & -- & -- \\
$L_3$ & Names only                 & 3NF & -- & -- & -- \\
$L_4$ & + types and primary keys   & 3NF & \ding{51} & -- & -- \\
$L_5$ & + foreign keys             & 3NF & \ding{51} & \ding{51} & -- \\
$L_6$ & + explicit join paths      & 3NF & \ding{51} & \ding{51} & \ding{51} \\
\bottomrule
\end{tabular}
\end{table}
The structural axis is discretised into six levels $L_1$ through $L_6$, ranging from a fully denormalised single table to a fully normalised schema augmented with rich relational metadata (Table~\ref{tab:structural_levels}). At the lowest level, $L_1$ executes against a materialised single-table 1NF database, with the prompt containing only that wide table and a denormalisation notice; $L_2$ behaves analogously but splits the schema into several wide clusters centred on fact or hub entities, accompanied by the same notice. \textbf{Crucially, $L_3$ through $L_6$ share the same underlying 3NF SQLite database}: these four levels are evaluated on identical data and differ only in the amount of structural metadata exposed in the prompt. Specifically, $L_3$ exposes only table and column names; $L_4$ adds column types and primary key / not-null constraints; $L_5$ adds foreign-key relationships annotated with cardinality; and $L_6$ adds explicit join-path descriptions specifying how tables can be connected. This design separates prompt-level structural exposure ($L_3$–$L_6$) from database-level structural choice ($L_1$/$L_2$): the former varies only the schema text, the latter varies the physical database itself. The materialisation of the 1NF and 2NF databases is described in §\ref{sec:denormalisation}; the additional EX-comparison considerations introduced by these denormalised levels are addressed in §\ref{sec:evaluation_protocol}.

\subsubsection{Semantic Axis}
\label{sec:semantic_axis}
\begin{table}[t]
\centering
\caption{Semantic levels ($S_1$--$S_3$), illustrated using example identifiers.}
\label{tab:semantic_levels}
\small
\setlength{\tabcolsep}{4pt}
\begin{tabularx}{\columnwidth}{clX}
\toprule
Level & Naming scheme & Example identifiers \\
\midrule
$S_1$ & Anonymous 
      & \texttt{col\_a}, \texttt{col\_b} \\
$S_2$ & Abbreviated 
      & \texttt{cust\_id}, \texttt{dist\_nm} \\
$S_3$ & Descriptive 
      & \texttt{customer\_id}, \texttt{district\_name} \\
\bottomrule
\end{tabularx}
\end{table}

The semantic axis is discretised into three levels $S_1$ through $S_3$, controlling the informativeness of schema identifiers without altering the underlying data (Table~\ref{tab:semantic_levels}). The manipulation applies only to the schema text in the prompt; column references in the generated SQL are mapped back to the original identifiers via a deterministic dictionary before execution, so the same gold queries can be used across all semantic levels. At the lowest level, $S_1$ replaces every table and column name with an anonymous positional identifier (\texttt{col\_a}, \texttt{col\_b}, \ldots), preserving the original column order within each table; we discuss the resulting ordering leakage in §\ref{sec:limitation}. $S_2$ uses abbreviated names a \textsc{Claude Code} agent that normalises mixed original naming conventions in BIRD into consistent short forms. $S_3$ uses fully descriptive English names derived from BIRD's column-description metadata, normalised to \texttt{lowercase\_snake\_case}. The mappings at each level are fixed across questions, ensuring that the same identifier rewrite is applied for every reference to a column; columns without a predefined mapping retain their original BIRD names across all semantic levels.

\subsection{Denormalisation Pipeline}
\label{sec:denormalisation}

\paragraph{Denormalisation targets.} BIRD provides only 3NF schemas; supporting $L_1$ and $L_2$ requires us to produce 1NF and 2NF databases for each of the nine target schemas. The $L_1$ target is a single wide table that fully flattens the relational structure, removing the need for any join at query time. The $L_2$ target is multiple wide clusters, each centred on a fact entity, preserving granularity where merging sibling facts would introduce ambiguity. For each database, we follow a hand-authored join plan that orders tables by key overlap, joining them through chained \texttt{FULL OUTER JOIN}, with \texttt{COALESCE} maintaining join keys across stages and table-prefixed column names to avoid collisions.

\paragraph{Engineering challenges.} This procedure faces four non-trivial challenges. First, a standard \texttt{LEFT JOIN} would discard orphan tuples that exist on only one side of a relationship; we therefore use chained \texttt{FULL OUTER JOIN}, so that both hub-only and dimension-only tuples are retained with missing attributes represented as \texttt{NULL}. Second, one-to-many and many-to-many relationships cause multiplicative row growth, producing intractable row counts on \texttt{card\_games} and \texttt{codebase\_community}, which we exclude (see §\ref{sec:dataset}). Third, \texttt{NULL} values introduced by full outer joins propagate through join keys and can block downstream joins; we mitigate this by propagating keys with \texttt{COALESCE} over previously joined attributes and ordering joins to attach high-overlap relationships before sparse dimensions. Fourth, chained full outer joins are not commutative in practice, different orderings yield different \texttt{NULL} patterns, so we specify a hand-authored join plan per database to make the denormalisation deterministic.

A key challenge arises for the first and second structural levels during evaluation. Denormalised schemas introduce duplicate rows due to join expansion, while the gold SQL queries are defined over the original normalised schema. Direct execution comparison is therefore not reliable without rewriting the gold queries into equivalent forms for each denormalised schema, which is infeasible to perform at scale. Rather than modifying evaluation, we treat duplicate handling as part of the task difficulty. Specifically, we include a fixed schema denormalisation notice in the prompt that instructs the model to account for redundancy using distinct clauses, count distinct aggregations, or deduplicating subqueries when necessary. This ensures consistent evaluation across all structural levels while reflecting the additional reasoning required under denormalised representations.

\paragraph{Residual information loss.} Despite these mitigations, the resulting 1NF wide tables remain lossy and redundant: \texttt{NULL} padding from outer joins and fan-out from multi-valued relationships mean that some queries solvable on the 3NF database have no exact equivalent on the 1NF table. We treat the residual impact on EX measurement as part of the task difficulty at $L_1$/$L_2$ rather than redesigning the metric, and quantify it in §\ref{sec:evaluation_protocol}.

\section{Evaluation Protocol}
\label{sec:evaluation_protocol}

\paragraph{Execution accuracy.} We adopt BIRD's execution accuracy (EX): a generated SQL is correct iff its result set matches that of the gold query. At $L_3$--$L_6$, predictions and gold queries execute on the same 3NF database, so EX is directly meaningful. At $L_1$ and $L_2$, however, predictions execute on the denormalised wide database while gold queries execute on the 3NF database, so result-set equivalence is non-trivial.

\paragraph{Validity risks and mitigation.} Two risks arise on the wide database: fan-out from one-to-many joins inflates aggregates (\texttt{SUM}, \texttt{AVG}, \texttt{COUNT}), and \texttt{NULL}-padded rows from full outer joins contaminate set membership. We \emph{partially} mitigate both by adding a fixed denormalisation notice to every $L_1$/$L_2$ prompt that instructs the model to use \texttt{DISTINCT}, \texttt{COUNT(DISTINCT ...)}, or deduplicating subqueries. 
\paragraph{Robustness.} Our main finding---asymmetric substitution between the two axes---rests on within-column comparisons (varying $S$ at fixed $L$), where both conditions share the same database and EX procedure, so any metric brittleness cancels out. Cross-$L$ comparisons at fixed $S$ should be interpreted with this caveat in mind.

\subsection{Experimental Setup}
\label{sec:setup}
\begin{table}[h]
\centering
\caption{Schema size and question count for the nine BIRD databases used in our study, on the original 3NF SQLite distribution. \emph{Avg c/t} denotes average columns per table. Schemas span a broad range of table count (3--13) and width (2.8--29.7 cols/table).}
\label{tab:benchmark_databases}
\small
\setlength{\tabcolsep}{6pt}
\begin{adjustbox}{max width=\columnwidth,center}
\begin{tabular}{lcccc}
\toprule
Database & Tables & Cols & Avg c/t & $n$ \\
\midrule
\texttt{california\_schools}      & 3  & 89  & 29.7 & 30  \\
\texttt{debit\_card\_specializing}& 5  & 21  & 4.2  & 30  \\
\texttt{european\_football\_2}    & 7  & 199 & 28.4 & 51  \\
\texttt{financial}                & 8  & 55  & 6.9  & 30  \\
\texttt{formula\_1}               & 13 & 94  & 7.2  & 66  \\
\texttt{student\_club}            & 8  & 48  & 6.0  & 48  \\
\texttt{superhero}                & 10 & 31  & 3.1  & 52  \\
\texttt{thrombosis\_prediction}   & 3  & 64  & 21.3 & 50  \\
\texttt{toxicology}               & 4  & 11  & 2.8  & 40  \\
\midrule
\textbf{Total}                    & \textbf{61} & \textbf{612} & --- & \textbf{397} \\
\bottomrule
\end{tabular}
\end{adjustbox}
\end{table}
\paragraph{Adopted dataset.}
Our evaluation uses the nine BIRD databases listed in Table~\ref{tab:benchmark_databases}, spanning a broad range of schema scale (3--13 tables, 11--199 columns, average 2.8--29.7 columns per table) and yielding $n=397$ questions from the corrected Arcwise-Plat-SQL subset (§\ref{sec:dataset}). The two databases excluded from the original 11 (\texttt{card\_games}, \texttt{codebase\_community}) cannot be materialised into a 1NF wide table within tractable row counts.

\paragraph{Models.} We evaluate ten instruction-tuned models spanning one closed and four open families: Gemini~2.5~Flash (Google, closed) \cite{comanici2025gemini}, accessed via the Google AI Studio API; Phi-4 (Microsoft, 14B) \cite{abdin2024phi} and OLMo-2-13B-Instruct (AllenAI, 13B) \cite{olmo20242}, run locally; and the full Qwen2.5-Coder-Instruct \cite{hui2024qwen2} size ladder (Alibaba) at 0.5B, 1.5B, 3B, 7B, 14B, and 32B parameters, also run locally. Local models are served in-process with Hugging Face \texttt{transformers} on a single GPU per model. No fine-tuning is applied; all models are queried using the chat-template instruction format shipped with each checkpoint.

\paragraph{Inference.} All models use greedy decoding with a single sample per query (\texttt{temperature}=0 for API providers and \texttt{do\_sample}=False for local \texttt{transformers}), without self-consistency, candidate selection, or post-hoc correction. Thus, the observed accuracy reflects the effect of schema representation rather than decoding-time search. We cap generation at $2{,}048$ new tokens for all open-model providers (OpenRouter and local \texttt{transformers}); for Gemini, we leave \texttt{max\_output\_tokens} at the SDK default. On transient or rate-limit (HTTP~429) errors, we retry up to five times with exponential backoff starting at 10\,s; queries that still fail return an empty string and are scored as incorrect.

\paragraph{Prompting.} Each prompt is constructed in a zero-shot setting and contains four components: (i) a fixed task instruction, (ii) the schema text rendered for the current $(L_i, S_j)$ condition (\S\ref{sec:schema_rep}), (iii) the natural language question, and (iv) for $L_1$ and $L_2$ only, a denormalisation notice. The full template is provided in Appendix~\ref{app:prompt}. No in-context examples are used in any condition, and identical prompts are sent to every model.

\paragraph{Execution and scale.} Generated SQL is executed on the corresponding database using SQLite~3.50.2 through the Python~3 \texttt{sqlite3} module, with a 120-second wall-clock timeout per query; timeouts are counted as incorrect. For $L_1$ and $L_2$, predictions are executed against the materialised denormalised SQLite files (\texttt{\{db\}\_\_1nf.sqlite} and \texttt{\{db\}\_\_2nf.sqlite}), while gold SQL runs on the native 3NF database. For $L_3$--$L_6$, both predicted and gold SQL run on the native 3NF database. Each (model, $L_i$, $S_j$) condition uses the same 397-question set, giving $10 \times 18 \times 397 = 71{,}460$ generations in total. We report execution accuracy as defined in \S\ref{sec:evaluation_protocol}.
\section{Experiments}

\subsection{Main Results}
\label{sec:main_results}
\paragraph{Overall Accuracy Across Schema Conditions.}
\begin{figure}[ht]
    \centering
    \includegraphics[width=\linewidth]{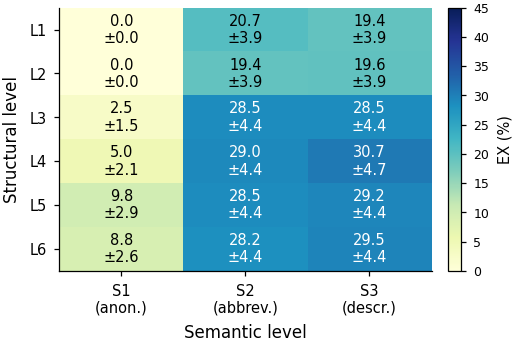}
    \caption{Marginal accuracy by axis across four models.}
    \label{fig:qwen_heatmap}
\end{figure}
We first characterise the full $(L_i, S_j)$ landscape before isolating either axis. Figure~\ref{fig:qwen_heatmap} reports execution accuracy across all 18 conditions for Qwen2.5-Coder-14B, and the heatmap separates into three regimes: the denormalised band ($L_1$, $L_2$) collapses to $0.0\%$ under $S_1$ and saturates at $19$--$21\%$ under $S_2$/$S_3$; the 3NF band ($L_3$--$L_6$) lifts this plateau to $28$--$31\%$; and the $S_1$ column stays an order of magnitude below $S_2$/$S_3$ at every structural level. Within $S_1$, accuracy rises monotonically from $0.0\%$ at $L_1$ to $9.8\%$ at $L_5$ before flattening, whereas within $S_2$/$S_3$ structural enrichment beyond $L_3$ adds little. The grid is therefore partitioned far more sharply by semantics than by structure, with structural gains concentrated where semantics is weakest.

\paragraph{Comparing the Two Axes.}
\begin{figure}[ht]
    \centering
    \includegraphics[width=\linewidth]{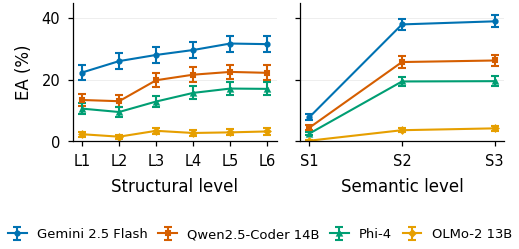}
    \caption{Marginal accuracy along each axis (pooled over the other) for four backbones. The semantic axis spans 2--3$\times$ the range of the structural axis, driven by a steep $S_1 \to S_2$ step; the structural sweep rises gradually and plateaus at $L_5$. Error bars: 95\% bootstrap CI.}

    \label{fig:two_axes}
\end{figure}
To quantify each axis and test whether the pattern generalises across backbones, Figure~\ref{fig:two_axes} collapses the grid into marginal sweeps for Gemini~2.5~Flash, Qwen2.5-Coder-14B, Phi-4, and OLMo-2-13B. The structural sweep rises gently and near-monotonically from $L_1$ to $L_5$ and plateaus at $L_5$--$L_6$, with a total $L_1\!\to\!L_6$ swing of $8$--$10$ points on the stronger backbones, while the semantic sweep is dominated by a steep $S_1\!\to\!S_2$ step and is essentially flat from $S_2$ to $S_3$. The semantic axis spans two to three times the range of the structural axis on every model tested, indicating that the asymmetry is a property of schema representation rather than of any single backbone.

\paragraph{Semantics Compensates for Structure, but Not Vice Versa.}

\begin{figure}[ht]
    \centering
\includegraphics[width=\linewidth]{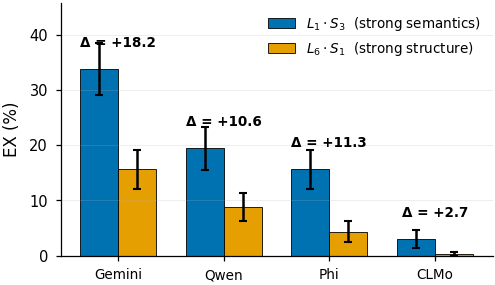}
    \caption{Diagonal corners $L_1\!\cdot\!S_3$ vs $L_6\!\cdot\!S_1$; gap positive on every model.}
    \label{fig:diagonal}
\end{figure}
We next test whether the two axes substitute for each other by pitting their extremes against each other. Figure~\ref{fig:diagonal} contrasts the two diagonal corners of the grid: $L_1\!\cdot\!S_3$ (weak structure, strong semantics) against $L_6\!\cdot\!S_1$ (strong structure, weak semantics). Across every model, $L_1\!\cdot\!S_3$ wins by a clear margin: $+18.2$ on Gemini, $+10.6$ on Qwen-14B, $+11.3$ on Phi-4, and $+2.7$ on OLMo, so a flattened single-table prompt with descriptive identifiers beats a fully normalised 3NF prompt with types, foreign keys, and explicit join paths under opaque identifiers. The substitution is one-directional: meaningful names let the model recover unexposed structure, but no amount of structural metadata, including explicit join paths at $L_6$, closes the gap when names provide no lexical anchor.
\paragraph{Effects Across Question Difficulty.}
Finally, we ask whether the two axes target the same questions or different bottlenecks. Table~\ref{tab:difficulty_gain} decomposes the marginal gains by BIRD difficulty tier, showing that almost all of $\Delta_S$ comes from the $S_1\!\to\!S_2$ step ($+42.3$ of $+43.0$ on simple; $+28.5$ of $+28.6$ on moderate) and that the total semantic gain shrinks with difficulty ($+43.0\!\to\!+21.8$). In contrast, the improvement from S2 to S3 is modest overall but more pronounced for challenging questions ($+3.3$), compared to minimal changes for simple and moderate questions ($+0.7$, $+0.1$). The structural axis is smaller but uniform across tiers ($\Delta_L \in [+8.8, +9.5]$), and the weight of $L_3\!\to\!L_6$ within $\Delta_L$ grows with difficulty ($+3.2$, $+2.6$, $+5.5$). The two axes thus resolve different bottlenecks, semantics rescues schema grounding and pays off most on easier questions, while structural metadata supports relational reasoning and becomes relatively more important as questions get harder, a split we trace to distinct failure modes in §\ref{sec:error_analysis}.

\begin{table}[t]
\centering
\caption{Per-difficulty accuracy gains along the semantic and structural axes for the primary model. Semantic gains are dominated by $S_1 \to S_2$ and decrease with difficulty, whereas structural gains are smaller and more uniform.}
\label{tab:difficulty_gain}
\small
\setlength{\tabcolsep}{5pt}
\begin{adjustbox}{max width=\columnwidth}
\begin{tabular}{lcccccc}
\toprule
\multirow{2}{*}{Difficulty} & \multicolumn{3}{c}{Semantic axis} & \multicolumn{3}{c}{Structural axis} \\
\cmidrule(lr){2-4} \cmidrule(lr){5-7}
 & $S_1{\to}S_2$ & $S_2{\to}S_3$ & Total $\Delta S$ & $L_1{\to}L_3$ & $L_3{\to}L_6$ & Total $\Delta L$ \\
\midrule
Simple      & +42.3 & +0.7 & +43.0 & +5.6 & +3.2 & +8.8 \\
Moderate    & +28.5 & +0.1 & +28.6 & +6.8 & +2.6 & +9.4 \\
Challenging & +18.5 & +3.3 & +21.8 & +4.0 & +5.5 & +9.5 \\
\bottomrule
\end{tabular}
\end{adjustbox}
\end{table}

\subsection{Additional Analysis}
\paragraph{Robustness across databases.}
\begin{figure}[ht]
    \centering
\includegraphics[width=\linewidth]{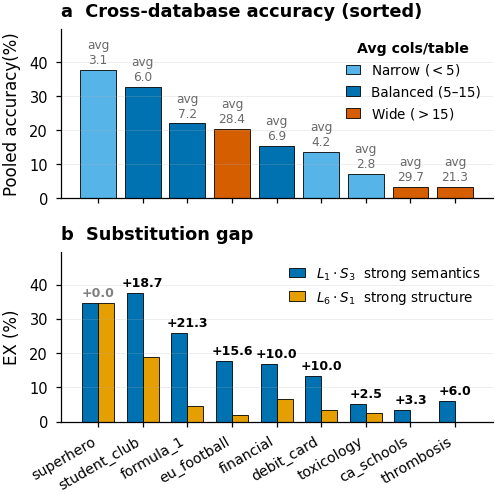}
    \caption{Per-database robustness (Qwen2.5-Coder 14B): (a) pooled accuracy by schema width; (b) substitution gap, positive in 8/9 databases.}
    \label{fig:database_robust}
\end{figure}
The pooled results in 
§\ref{sec:main_results} aggregate over 9 BIRD databases
of substantially different shape, so we verify on Qwen2.5-Coder 14B finding is not driven by any single database.
Figure~\ref{fig:database_robust} stacks two views along the same database
ordering, sorted by pooled accuracy. The top panel shows that pooled
accuracy across the 18 schema conditions varies more than tenfold across
databases (3.3\% to 37.7\%), with the dominant explanatory variable being
schema \emph{width}: the three top-ranked databases all have narrow or
balanced tables (avg 3.1--7.2 cols/table), while two of the three
bottom-ranked databases are wide (avg $>20$). Notably, two narrow-schema
databases (\texttt{toxicology}, \texttt{debit\_card}) also fall near the
bottom, indicating that narrow tables alone are necessary but not sufficient
for high accuracy, schema typology interacts with question difficulty.
Reading vertically, the bottom panel reports the substitution gap
$L_1\cdot S_3 - L_6\cdot S_1$ per database in the same column order. The
gap is positive in 8 of 9 databases with a single tie on \texttt{superhero}
, where narrow tables and rich relational metadata are sufficient even
under anonymous identifiers, suggesting a ceiling effect rather than a
counterexample, and no negative cases. Across the 9 databases the gap
spans from 0 to +21.3 pp, with the largest gaps occurring on mid-range
databases (\texttt{formula\_1} +21.3, \texttt{student\_club} +18.7) where
both corner cells have headroom to differ. The asymmetric semantic
substitution therefore reproduces within individual databases, not only in
the pooled average, and is robust to schema-typology variation across BIRD.

\paragraph{Does model capacity change the picture?}
\begin{figure}[ht]
    \centering
\includegraphics[width=\linewidth]{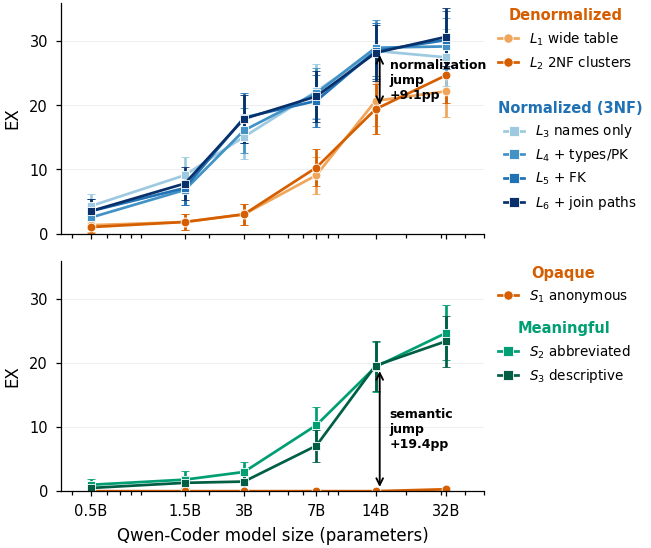}
    \caption{Qwen2.5-Coder scaling (0.5B--32B). Top: structural axis at $S_2$. Bottom: semantic axis at $L_2$. Normalisation jump emerges at 3B; semantic jump emerges later (7B) but grows larger with scale.}
    \label{fig:5_3_3_scaling}
\end{figure}
Sweeping the Qwen2.5-Coder family across six sizes (0.5B--32B) at production-typical baselines ($S_2$ when varying the structural axis, $L_2$ when varying the semantic axis), we find that both axes scale as step functions rather than smoothly (Fig.~\ref{fig:5_3_3_scaling}): the normalisation jump $L_2 \to L_3$ emerges at 3B and stays roughly constant (+9--12\,pp), while the meaningfulness jump $S_1 \to S_2$ emerges only at 7B but climbs to +24.4\,pp at 32B---more than twice the structural jump at scale; below 3B neither axis delivers a usable signal, so we read the semantic--structural asymmetry as an emergent property of capable text-to-SQL systems rather than a universal one.
\subsection{Error Analysis: How Schema Shapes Failure Modes}
\label{sec:error_analysis}

\paragraph{Error Categories and Annotation Procedure.}
We classify every execution-accuracy failure into five mutually exclusive categories: \emph{weak deduplication} (aggregates that ignore fan-out from denormalisation), \emph{wrong column}, \emph{wrong table}, \emph{join plan} (incorrect keys or omitted relations), and \emph{logic-other} (filter, literal, and aggregate-arithmetic errors largely independent of schema structure). Classification is delegated to an agentic \textsc{Claude Code}\footnote{\url{https://www.anthropic.com/claude-code}} pipeline driven by Claude~Sonnet~4.5: for each failure the agent receives the question, gold and predicted SQL, both result sets, and the schema text for that condition, and assigns one category with a short justification. 

\begin{table}[t]
\centering
\caption{Distribution of error categories. Left: varying $L$ at fixed $S_3$. Right: varying $S$ at fixed $L_3$. Cells are percentages; columns sum to 100\% up to rounding. ``--'' marks categories impossible by construction. $n$ is the number of failures. Reported for Gemini~2.5~Flash.}
\label{tab:error_distribution}
\small
\setlength{\tabcolsep}{3pt}
\begin{adjustbox}{max width=\columnwidth}
\begin{tabular}{l|cccccc|ccc}
\toprule
& \multicolumn{6}{c|}{$S_3$ fixed, varying $L$} & \multicolumn{3}{c}{$L_3$ fixed, varying $S$} \\
Error type & $L_1$ & $L_2$ & $L_3$ & $L_4$ & $L_5$ & $L_6$ & $S_1$ & $S_2$ & $S_3$ \\
\midrule
Weak deduplication  & 32 & 29 &  1 &  1 &  1 &  1 &  0 &  1 &  1 \\
Wrong column        &  2 &  5 &  1 &  2 &  1 &  2 & 78 & 14 &  1 \\
Wrong table         &  -- & -- & 31 & 29 & 30 & 28 &  8 & 25 & 31 \\
Join plan           &  -- &  4 & 11 & 12 & 10 & 11 &  2 & 12 & 11 \\
Logic-other         & 66 & 62 & 56 & 56 & 58 & 58 & 12 & 48 & 56 \\
\midrule
$n$ (failures)      & 263 & 242 & 235 & 236 & 237 & 239 & 379 & 244 & 235 \\
\bottomrule
\end{tabular}
\end{adjustbox}
\end{table}

\paragraph{Structure Redistributes Errors; Opaque Names Collapse Them.}
Structural enrichment redistributes errors rather than reducing them (Table~\ref{tab:error_distribution}, left). At $L_1$, $32\%$ of failures are weak-deduplication errors from join fan-out (Table~\ref{tab:error_cases}a) despite the \texttt{DISTINCT} hint in our denormalisation notice, but the same questions almost never fail this way at $L_3$ ($1\%$); the error mass reappears as \emph{wrong table} ($31\%$, Table~\ref{tab:error_cases}c) and \emph{join plan} ($11\%$), both absent at $L_1$ by construction. This composition is then stable across $L_3$--$L_6$ (wrong-table within $3$\,pp, join-plan in the $10$--$12\%$ band), mirroring the flat $L_3 \to L_6$ accuracy plateau in §5.1: even at $L_5$, where foreign keys are explicit, the model still joins on canonical key names rather than BIRD-specific link columns (Table~\ref{tab:error_cases}d), direct evidence that extra structural metadata does not absorb into the model's join choices. The semantic axis (right panel) shows the dual collapse: under $S_1$ at $L_3$, \emph{wrong column} alone accounts for $78\%$ of failures because anonymous identifiers leave no lexical signal for grounding question terms; as semantics strengthens, this single mode splits into the structural categories visible at $S_3$. The residual \emph{logic-other} share is near-constant across both axes, exemplified by arithmetic errors that recur verbatim on 1NF and 3NF predictions (Table~\ref{tab:error_cases}e) --- the kind of failure on which schema representation has no leverage.

\begin{table}[t]
\centering
\caption{Representative SQL failure cases. \textcolor{red}{Red} marks predicted (wrong) tokens; \textcolor{blue}{blue} marks gold (correct) tokens. (a) denormalised-side, (b)--(c) normalised-side.}
\label{tab:error_cases}
\small
\begin{tabular}{p{0.95\columnwidth}}
\toprule

\textbf{(a) Weak deduplication} \hfill $L_1 \cdot S_3$ \\
\textit{Q: How many female clients are there? (\texttt{financial})}
\begin{tcolorbox}
SELECT COUNT(\textcolor{blue}{\checkmark DISTINCT}~client\_client\_id) \\
FROM one\_nf\_0 WHERE client\_gender = 'F';
\end{tcolorbox}
\emph{Wide table replicates each client per transaction; \texttt{DISTINCT} hint ignored.}\\[2pt]

\noindent\rule{\columnwidth}{0.3pt}\\[-2pt]

\textbf{(b) Wrong table} \hfill $L_3 \cdot S_3$ \\
\textit{Q: Notes for the event on 2019-09-14? (\texttt{student\_club})}
\begin{tcolorbox}
SELECT notes FROM \textcolor{red}{income} / \textcolor{blue}{event} \\
WHERE event\_date = '2019-09-14';
\end{tcolorbox}
\emph{Both \texttt{event} and \texttt{income} carry \texttt{notes} at 3NF; wrong base table.}\\[2pt]

\textbf{(c) Join plan despite explicit foreign keys} \hfill $L_5 \cdot S_3$ \\
\textit{Q: Which members attended the AGM event? (\texttt{student\_club})}
\begin{tcolorbox}
SELECT ... FROM member AS T1 \\
INNER JOIN attendance AS T2 \\
\hspace*{1em} ON T1.member\_id = \\
\hspace*{2em} \textcolor{red}{T2.member\_id} / \textcolor{blue}{T2.link\_to\_member} \\
INNER JOIN event AS T3 \\
\hspace*{1em} ON \textcolor{red}{T2.event\_id} / \textcolor{blue}{T2.link\_to\_event} \\
\hspace*{2em} = T3.event\_id;
\end{tcolorbox}
\emph{FKs explicit at $L_5$; model still defaults to canonical key names.}\\

\bottomrule
\end{tabular}
\end{table}

\section{Conclusion}
We studied how schema representation shapes LLM-based text-to-SQL by crossing structural observability ($L_1$--$L_6$) with semantic informativeness ($S_1$--$S_3$) in a controlled 18-condition factorial, and found an asymmetric substitution between the two axes: meaningful names compensate for missing structure while richer structural metadata does not recover performance under opaque names. This asymmetry reproduces within 8 of 9 BIRD databases and emerges with model scale (negligible below 3B), and within the structural axis the gain is concentrated at the normalisation step $L_2 \to L_3$ rather than in metadata layered on top of 3NF---suggesting that for current prompt-based text-to-SQL the practical lever is column naming and basic normalisation, not elaborate schema annotation.

\section*{Limitations}
\label{sec:limitation}

Our study has four main limitations. First, the factorial design is evaluated on a single benchmark family (a corrected BIRD mini-dev subset \citep{jin2026pervasive}) and further restricted to nine databases for which 1NF/2NF materialisation via chained \texttt{FULL OUTER JOIN} is computationally tractable; whether the asymmetric substitution effect generalises to denser many-to-many schemas or to non-BIRD benchmarks is untested. Second, at $L_1$ and $L_2$ predictions execute on the denormalised wide database while gold SQL executes on the 3NF database, so result-set equivalence is not guaranteed; we therefore read $L_1$/$L_2$ accuracy as a lower bound and ground the main finding on within-$L$ comparisons where this metric brittleness cancels out. Third, $S_1$ replaces identifiers with positional names (\texttt{col\_a}, \texttt{col\_b}, \ldots) while preserving original column order, leaking a weak structural signal (e.g.\ keys typically appearing first); our $S_1$ results therefore upper-bound performance under a fully content-free naming scheme. 


\bibliography{custom}

\appendix

\section{Prompt Details}
\label{app:prompt_details}
\subsection{Text-to-SQL prompt}

\begin{tcolorbox}[
    colback=gray!5, 
    colframe=gray!20, 
    arc=4pt, 
    boxrule=0.5pt, 
    fontupper=\small\ttfamily,
    left=8pt, right=8pt, top=8pt, bottom=8pt
]
You are an expert SQLite assistant. Given the database schema and the question below, write a single SQLite SELECT query that correctly answers the question.

Rules: \\
- Output the SQL query only — no explanation, no markdown, no code fences. \\
- Use only the tables and columns defined in the schema.\\
- Do not invent column or table names.\\

Schema: \{schema\} \\
Question: \{question\} \\
SQL:
\end{tcolorbox}

\subsection{Denormalization Notice}
\label{app:denormalization_notice}

\begin{tcolorbox}[
    colback=gray!5,
    colframe=gray!20,
    arc=4pt,
    boxrule=0.5pt,
    breakable,
    left=8pt, right=8pt, top=8pt, bottom=8pt,
    nobeforeafter
]
\begin{lstlisting}[
    breaklines=true,
    breakindent=0pt,
    breakautoindent=false,
    basicstyle=\small\ttfamily,
    columns=flexible,
    keepspaces=true,
    showstringspaces=false,
    aboveskip=0pt,
    belowskip=0pt,
    xleftmargin=0pt,
    framexleftmargin=0pt,
]
Schema Denormalization Notice

The database schema provided is in a denormalized form. Tables have been constructed by joining multiple normalized entities together, which means a single logical record (e.g. one race result, one driver) may appear across multiple rows due to join redundancy. You must account for this when writing SQL to avoid returning inflated or incorrect results.

The rules below tell you exactly how to handle this for each query type.

---

1. Retrieval Queries (SELECT without aggregation)

Retrieval queries return rows of values - names, IDs, descriptions, dates.

RULE: Always use SELECT DISTINCT when retrieving any column or combination of columns, unless the question explicitly asks for all occurrences including repetitions.

Examples:
  Question: "List the names of all drivers who competed in race 5."
  Wrong:   SELECT driver_name FROM results_flat WHERE race_id = 5
  Correct: SELECT DISTINCT driver_name FROM results_flat WHERE race_id = 5

  Question: "What circuits have hosted a race?"
  Wrong:   SELECT circuit_name FROM results_flat
  Correct: SELECT DISTINCT circuit_name FROM results_flat

When to NOT use DISTINCT on retrieval:
  - The question asks for all rows explicitly, e.g. "list every lap time recorded" - in this case repetition may be expected.
  - The question includes an ORDER BY with LIMIT and you are retrieving a specific ranked row - DISTINCT may change ranking behaviour.

---

2. COUNT Queries

COUNT queries count how many things exist. This is the most error-prone query type on denormalized schemas. You must identify what entity is being counted and apply DISTINCT to that entity's primary identifier.

RULE: Always use COUNT(DISTINCT <entity_id>) rather than COUNT(*) or COUNT(<column>), where <entity_id> is the primary key of the logical entity the question is asking about.

Examples:
  Question: "How many drivers competed in race 5?"
  Wrong:   SELECT COUNT(*) FROM results_flat WHERE race_id = 5
  Wrong:   SELECT COUNT(driver_id) FROM results_flat WHERE race_id = 5
  Correct: SELECT COUNT(DISTINCT driver_id) FROM results_flat WHERE race_id = 5

  Question: "How many races has Hamilton participated in?"
  Wrong:   SELECT COUNT(*) FROM results_flat WHERE driver_name = 'Hamilton'
  Correct: SELECT COUNT(DISTINCT race_id) FROM results_flat WHERE driver_name = 'Hamilton'

  Question: "How many results were recorded with more than 5 points?"
  Correct: SELECT COUNT(DISTINCT result_id) FROM results_flat WHERE points > 5

Identifying which entity_id to use:
  - If counting people/drivers/teams -> DISTINCT on driver_id or team_id
  - If counting events/races/games -> DISTINCT on race_id or event_id
  - If counting records/results -> DISTINCT on result_id
  - If counting laps -> lap rows are the atomic unit; COUNT(*) may be appropriate if laps themselves are what is counted

---

3. SUM and AVG Queries

SUM and AVG are the most dangerous query types on denormalized schemas. A simple SUM(points) will add up points once per row, but each logical result may appear many times (once per lap), inflating the total massively. DISTINCT cannot be applied directly inside SUM or AVG.

RULE: Always deduplicate using a subquery before applying SUM or AVG. The subquery should SELECT DISTINCT on the primary key of the entity whose attribute you are aggregating, along with the attribute column.

Template:
  SELECT SUM(col) / AVG(col)
  FROM (
    SELECT DISTINCT <entity_id>, <col>
    FROM <table>
    WHERE <condition>
  )

Examples:
  Question: "What is the total points scored by Hamilton?"
  Wrong:   SELECT SUM(points) FROM results_flat WHERE driver_name = 'Hamilton'
           -- Inflated: sums points once per lap row

  Correct: SELECT SUM(points) FROM (
             SELECT DISTINCT result_id, points FROM results_flat WHERE driver_name = 'Hamilton')

  Question: "What is the average points per race in the 2020 season?"
  Correct: SELECT AVG(points) FROM (
             SELECT DISTINCT result_id, points FROM results_flat WHERE race_year = 2020)
---
4. GROUP BY Queries

GROUP BY queries aggregate within groups - e.g. total points per driver, number of races per year. On denormalized schemas, GROUP BY alone does not deduplicate - it groups all rows including duplicates.

RULE: When using GROUP BY, either:
  (a) Apply COUNT(DISTINCT <entity_id>) inside the aggregation, or
  (b) Deduplicate with a subquery before grouping

Examples:
  Question: "How many races did each driver compete in?"
  Wrong:   SELECT driver_name, COUNT(*) FROM results_flat GROUP BY driver_name
  Correct: SELECT driver_name, COUNT(DISTINCT race_id) FROM results_flat GROUP BY driver_name

  Question: "What is the total points per driver in the 2020 season?"
  Wrong:   SELECT driver_name, SUM(points) FROM results_flat WHERE race_year = 2020 GROUP BY driver_name
  Correct: SELECT driver_name, SUM(points) FROM (
             SELECT DISTINCT result_id, driver_name, points FROM results_flat WHERE race_year = 2020) 
             GROUP BY driver_name

\end{lstlisting}
\end{tcolorbox}

\subsection{S3 to S2 Mapping Prompt}
\begin{tcolorbox}[
    colback=gray!5, 
    colframe=gray!20, 
    arc=4pt, 
    boxrule=0.5pt, 
    fontupper=\small\ttfamily,
    left=8pt, right=8pt, top=8pt, bottom=8pt
]
You are an expert data engineer and schema designer. I need you to generate a mapping rule based on specific conditions, focusing on converting descriptive column names into standardized, abbreviated column names. \\
- If the input configuration is set to S3, ensure any incoming descriptive column names are converted to their S2 abbreviated equivalents. \\
- If the input configuration is set to S2, ensure any incoming abbreviated column names are converted to their S3 descriptive column names, using column description and value description in database description file.
\end{tcolorbox}

\label{app:prompt}

This is an appendix.

\end{document}